\documentstyle[prl,floats,aps,twocolumn,epsf,graphicx]{revtex}
\begin{document}
\twocolumn[\hsize\textwidth\columnwidth\hsize\csname
@twocolumnfalse\endcsname

\title{Wave Tails in Time Dependent Backgrounds}
\author{Shahar Hod}
\address{Department of Condensed Matter Physics, Weizmann Institute, 
Rehovot 76100, Israel}
\date{\today}

\maketitle

\begin{abstract}

\ \ \ It is well-known that waves propagating under the influence of 
a scattering potential develop ``tails''. 
However, the study of late-time tails has so far been restricted to 
time-independent backgrounds. In this paper we explore the late-time 
evolution of spherical waves propagating under the influence of a 
{\it time-dependent} scattering potential. 
It is shown that the tail structure is modified due to the temporal dependence of the 
potential. The analytical results are confirmed by numerical calculations.

\end{abstract}
\bigskip

]

The phenomenon of wave tails have fascinated many physicists and
mathematicians from the early explorations of wave theories. Wave
tails have found various applications from the first studies in light
propagation \cite{Huy} to the theory behind the proposed experiments to
detect gravitational waves \cite{BlanDam,BlanSch,Wise}. In fact, tail-free propagation seems to be the exception
rather than the rule \cite{Fried,Noon}. For instance, it is well established that
scalar, electromagnetic and gravitational waves in curved spacetimes 
propagate not only along light cones, but also spread
inside them. This implies that waves do not cut 
off sharply after the passage of the wave front, 
but rather leave a tail or wake at late times.

From a physical point of view, the most interesting mechanism for the
production of late-time tails is the backscattering of waves off a 
potential (or a spacetime curvature) at asymptotically far
regions \cite{Thorne,Price}. 
This can be described as follows. Consider a wave from a source point $y$. 
The late-time tail observed at a fixed spatial location, $x$, and at
time $t$, is a
consequence of the wave first propagating to a 
distant point $x' >>y,x$, being scattered by $V(x',t')$ at 
time $t' \simeq t/2$, and then returning
to $x$ at a time $t \simeq (x'-y) + (x'-x) \simeq 2x'$ \cite{Ching}. Hence, the
scattering amplitude (and thus the late-time tail itself) are expected
to be proportional to $V(x',t') \simeq V(t/2,t/2)$ (However, in a 
previous paper \cite{Hod0} we have shown that this picture is somewhat 
naive, and requires some important modifications.)

The propagation of spherical waves in curved spacetimes or in 
optical cavities is often governed by the 
Klein-Gordon (KG) equation \cite{Chan}

\begin{equation}\label{Eq1}
\left[ {{\partial ^2} \over {\partial t^2}}-{{\partial ^2} \over
    {\partial x^2}} +{1 \over {x^2_s}}V(x,t) \right]\Psi=0\  ,
\end{equation}
where $V(x,t)$ is an effective
curvature potential which determines the scattering of the waves by the
background geometry (we henceforth take $x_s=1$ without loss of
generality). It was first demonstrated by Price \cite{Price} 
that a (nearly spherical) collapsing
star leaves behind it a ``tail'' which decays asymptotically 
as an inverse power of time.

The analysis of Price has been extended by many authors. 
Gundlach, Price, and Pullin \cite{Gundlach2} showed that 
power-law tails are a genuine feature of gravitational collapse --
the existence of these tails was demonstrated in full {\it non}-linear 
numerical simulations of the collapse 
of a self-gravitating scalar 
field (this was later reproduced in \cite{BurOr}). 
Moreover, since the
late-time tail is a direct consequence of the scattering of the waves at 
asymptotically far regions, it has been pointed out that the same power-law tails
would develop independently of the existence of an 
horizon \cite{Gundlach1}. This implies that tails should also be
formed when the collapse fails to produce a black hole, or even in 
the context of stellar dynamics (e.g., in perturbations of neutron
stars). In recent years there is a flurry of activity in the field of 
wave tails, see e.g.,
\cite{Bicak,Leaver,SuPr,BomSeb,Andersson1,Andersson2,Nolan,Brad1,Brad2,HodPir123,HodPir4,Barack1,Hod1,KokSch,Krivan1,Krivan2,Ori1,Barack2,Hod2,BarOr,CaWa,Hod3,Hod4,Barack3,Krivan3,BrChLaPo,AnderKos,LaaPoi,Mal,WaAbMa,WaMeAb,MoRo,KoTo},
and references therein. 

Yet, in spite of the numerous works addressing the problem of wave
tails, a thorough understanding of this fascinating phenomenon 
is not complete. In particular, most
of previous analyses are restricted to the specific class of (time
independent) ``logarithmic
potentials'' of the form $V(x) \sim \ln^{\beta}x/x^{\alpha}$ 
(where $\alpha >2$ and $\beta=0,1$ are parameters) \cite{Ching}. 
Recently, we have given a systematic analysis of the tail phenomenon for 
waves propagating under the influence of 
a {\it general} time-independent scattering potential \cite{Hod0}.

It should be realized, however, that 
a realistic gravitational collapse produces a {\it time-dependent}
spacetime geometry, on which the tails are developing. 
This fact calls for a systematic exploration of the general
properties of wave tails in {\it dynamical} (time-dependent)
backgrounds. This is the aim of the present paper, in which
we present our main results.

We consider the evolution of a wave field whose dynamics is governed 
by a KG-type equation $\Phi_{;\nu}^{;\nu} +V(r,t)\Phi=0$. Substituting 
$\Phi =\Psi(t,r)/r$ ($r$ being the circumferential radius), 
one obtains a wave equation
of the form Eq. (\ref{Eq1}) \cite{Note}.

It proofs useful to introduce the double-null coordinates $u\equiv t-x$ 
and $v \equiv t+x$, which are a retarded time coordinate and an advanced
time coordinate, respectively. 
The initial data is in the form of some compact outgoing
pulse in the range $u_0 \leq u \leq u_1$, 
specified on an ingoing null surface $v=v_0$.

The general solution to the wave-equation (\ref{Eq1}) can be written
as a series depending on two arbitrary functions $F$ and $G$ \cite{Price}

\begin{eqnarray}\label{Eq2}
\Psi& = & {{G^{(0)}(u)+
F^{(0)}(v)}}\nonumber \\
 &&+  \sum\limits_{k=0}^\infty  {\left[B_k(u,v)
     G^{(-k-1)}(u) + C_k(u,v) F^{(-k-1)}(v)\right]}\  , \nonumber \\
\end{eqnarray}
For any function $H$, $H^{(k)}$ is its $k$th
derivative; negative-order derivatives are to be interpreted as
integrals [we shall also denote $\partial_u^m \partial_v^n H$ by
$H^{(m,n)}$.] The functions $B_k(u,v)$ satisfy the recursion relation 

\begin{equation}\label{Eq3}
{B_k}_{,v} = -{B_{k-1}}_{,uv}-{1 \over 4} V B_{k-1} \  ,
\end{equation}

%\begin{equation}\label{Eq3}
%B_k' + \dot B_k={1 \over 2}\left[B''_{k-1}-\ddot B_{k-1}- 
%VB_{k-1} \right]\  ,
%\end{equation}
for $k \geq 1$, 
%where $B'\equiv dB/dv$, $\dot B \equiv dB/dt$, 
and 

\begin{equation}\label{Eq4}
B_{0,v}=-V/4\  .
\end{equation}
%\begin{equation}\label{Eq4}
%B'_0+\dot B_0=-V(x,t)/4\  . 
%\end{equation}
%which implies 
%
%\begin{equation}\label{Eq4}
%B_0(u,v) =-{1 \over 4} \int \limits^v V(u,v') dv'\  .
%%V^{(-1)}/2\  , 
%\end{equation}

For the first Born approximation to be valid the scattering potential
$V$ should approach zero faster than $1/v^2$ as $v \to
\infty$, see e.g., \cite{Ching,HodPir123}. 
Otherwise, the scattering potential cannot be neglected at
asymptotically far regions [see Eq. (\ref{Eq6}) below]. 
The recursion relation, Eq. (\ref{Eq3}), yields
$B_k(u,v)=(-1)^{k+1} V^{(k,-1)}/4$.

It is useful to classify the
scattering potentials into two groups, according to their asymptotic
behavior:

\begin{itemize}
\item Group I: $|V_{,u}|$ approaches
  zero {\it faster} than $|V|$ as $v \to \infty$.
\item Group II: $|V_{,u}|$ approaches zero at the {\it same} rate as $|V|$ as $
v \to \infty$.
\end{itemize}

{\it Group I}. --- 
The first stage of the evolution is the scattering
of the field in the region $u_0 \leq u \leq u_1$. 
The first sum in Eq. (\ref{Eq2}) represents the primary waves in the
wave front (i.e., the zeroth-order solution, with $V \equiv 0$), 
while the second sum represents backscattered waves. The interpretation of
these integral terms as backscatter comes from the fact that they depend on
data spread out over a {\it section} of the past light cone, while outgoing
waves depend only on data at a fixed $u$ \cite{Price}.

After the passage of the primary waves there is no outgoing radiation 
for $u > u_1$, aside from backscattered waves. This means that $G(u_1) = 0$. 
Hence, at $u = u_1$ and for $v >> u_1$ (where $t \simeq x \simeq v/2$), 
the dominant term in Eq. (\ref{Eq2}) is 

\begin{equation}\label{Eq5}
\Psi(u=u_1,v) =B_0(u=u_1,v)G^{(-1)}(u_1)\  .
\end{equation}
This is the dominant backscatter of the primary waves.

With this specification of characteristic data on $u=u_1$, we shall
next consider the asymptotic evolution of the field. We confine our
attention to the region $u>u_1$, $x \gg x_s$. 
To a {\it first} Born approximation, the spacetime in this region is
approximated as flat \cite{Price,Gundlach1}. 
Thus, to first order in $V$ (that is, in a first Born approximation) 
the solution for $\Psi$ can be written as

\begin{equation}\label{Eq6}
\Psi ={g^{(0)}(u)+f^{(0)}(v)}\  .
\end{equation}
Comparing Eq. (\ref{Eq6}) with the initial data on $u=u_1$, 
Eq. (\ref{Eq5}), one finds

\begin{equation}\label{Eq7}
f(v)=-G^{(-1)}(u_1) V^{(0,-1)}(u=u_1, v)/4\  .
\end{equation}

For late times $t \gg x$ one can expand
$g(u)=\sum\limits_{n=0}^{\infty} (-1)^ng^{(n)}(t)x^n/n!$ and similarly
for $f(v)$. With these expansions, Eq. (\ref{Eq6}) can be rewritten as

\begin{equation}\label{Eq8}
\Psi =\sum\limits_{n=0}^{\infty}  {K_0^nx^n\left[ f^{(n)}(t)+(-1)^ng^{(n)}(t) \right]}\  ,
\end{equation}
where the coefficients $K_0^n$ are those given in \cite{Price}.

Using the boundary conditions for small $r$ [regularity as $x \to -\infty$, at
the horizon of a black hole, or at $x=0$, for a non-singular 
model (e.g., a stellar model)], one
finds that at late times $g(t)=-f(t)$ to first order in the 
scattering potential $V$ (see e.g., \cite{Price,Gundlach1} for
additional details). 
That is, the incoming and outgoing parts of
the tail are equal in magnitude at late-times. This almost total
reflection of the ingoing waves at small $r$ can easily be understood
on physical grounds -- it 
simply manifests the impenetrability of the barrier to low-frequency
waves \cite{Price} (which are the ones to dominate the late-time 
evolution \cite{Leaver}). 
We therefore find that the late-time behavior 
of the field at a fixed radius ($x \ll t$) 
is dominated by [see Eq. (\ref{Eq8})]

\begin{equation}\label{Eq9}
\Psi \simeq 2K_0^{1}f^{(1)}(t)x\  , 
\end{equation}
which implies

\begin{equation}\label{Eq10}
\Psi \simeq -2^{-1}K_0^{1} G^{(-1)}(u_1)x V(u=u_1,v=t)\  ,
\end{equation}
or equivalently

\begin{equation}\label{Eq11}
\Psi(x,t) \simeq -2^{-1}K_0^{1}G^{(-1)}(u_1)xV(t/2,t/2)\  .
\end{equation}

{\it Group II}. --- 
The dominant backscatter of the primary waves is 
$\Psi(u=u_1,x)=\sum\limits_{k=0}^{\infty} 
B_k(u=u_1,v)G^{(-k-1)}(u_1)$. 
Using an analysis along the same lines as before, one finds 

\begin{eqnarray}\label{Eq12}
\Psi& \simeq & 
\sum\limits_{n=1,3,...}^{\infty} 2^{-1}K_0^n x^n 
\sum\limits_{k=0}^{\infty} (-1)^{k+1} \nonumber \\
&& \times G^{(-k-1)}(u_1)V^{(k,n-1)}(u=u_1,v=t)\  ,
\end{eqnarray}
at late-times. 
Note that Eq. (\ref{Eq12}) is merely a generalization of
Eq. (\ref{Eq10}), and reduces to it if $|V_{,x}|$ or $|V_{,t}|$ approach 
zero faster than $|V|$ [in which case $V^{(0,0)}$ dominates at late-times].

{\it Numerical calculations}. --- 
It is straightforward to integrate Eq. (\ref{Eq1}) using the methods 
described in \cite{Gundlach1,Hod1}. 
The {\it late-time} evolution of the 
field is independent of the form of the initial data used. The
results presented here are for a Gaussian pulse. 

\begin{figure}[tbh]
\centerline{\epsfxsize=9cm \epsfbox{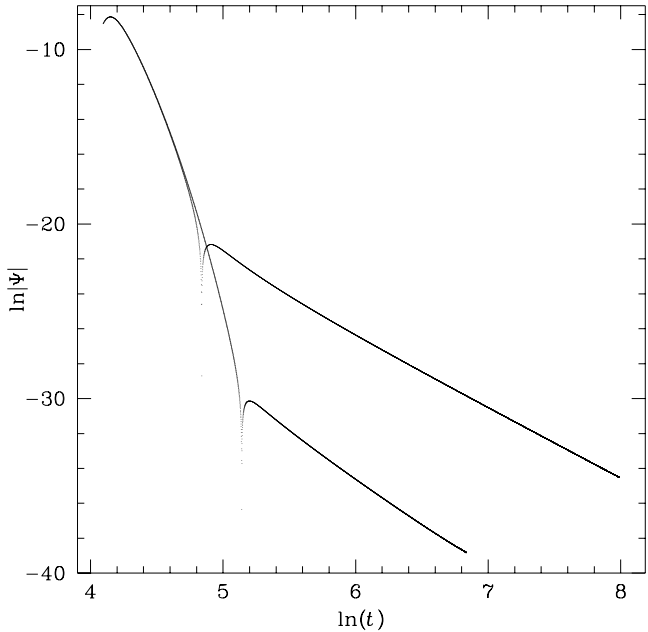}} 
\caption{Temporal evolution of the field for time-dependent 
scattering potentials of the form $V(x,t)=1/x^{\alpha}t^{\beta}$ 
(the results presented here are for $\alpha=3$.) The power-law indices 
are $-4.04$, and $-5.08$ for $\beta=1$ (upper graph), and $\beta=2$,
respectively. These values should be compared 
with the {\it analytically} predicted values of 
$-4,$ and $-5$, respectively.}
\label{Fig1}
\end{figure}

\begin{figure}[tbh]
\centerline{\epsfxsize=9cm \epsfbox{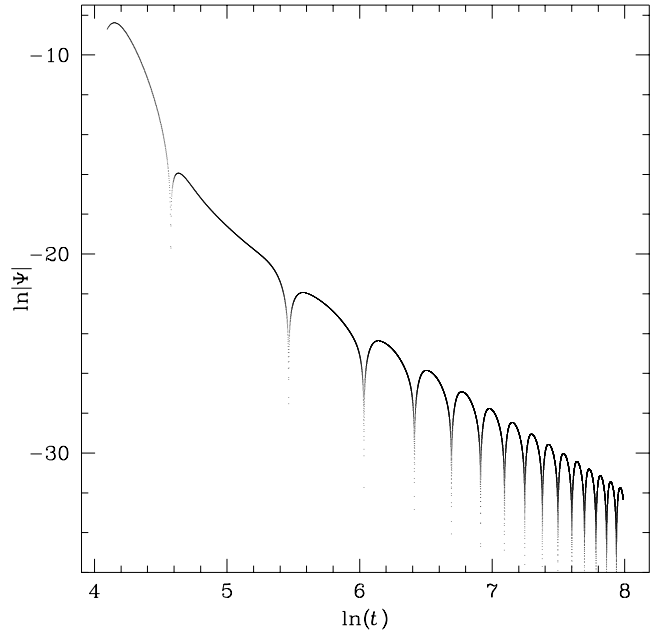}} 
\caption{Temporal evolution of the field for time-dependent 
scattering potentials of the form $V(x,t)=sin(\omega t)/x^{\alpha}$ 
(the results presented here are for $\alpha=4$, and $\omega=\pi/100$.) 
The slope (determined from the maxima of the oscillations) is $-4.07$, 
in excellent agreement with the {\it analytically} predicted value of $-4$. The 
frequency of the oscillations is $\omega$ to within $1\%$.}
\label{Fig2}
\end{figure}

The temporal evolutions of the waves 
(under the influence of the various scattering
potentials) are shown in Figs. \ref{Fig1} and \ref{Fig2} 
(We have studied other potentials as well, which are not shown
here.) We find an excellent agreement between the {\it analytical} results 
and the {\it numerical} calculations. 

In summary, we have explored the tail phenomena for 
spherical waves propagating under the influence of 
a general {\it time-dependent} scattering potential. 
It was shown that the late-time tail at a fixed spatial location 
is governed by the scattering potential itself, and by its derivatives 
(both the spatial and the temporal ones). The analytical results are in 
agreement with numerical calculations.

We are at present extending the analysis to 
include scattering
potentials that lack spherical symmetry (in which case the scattering
problem is of $2+1$ dimensions).

\bigskip
\noindent
{\bf ACKNOWLEDGMENTS}
\bigskip

I thank Tsvi Piran for discussions. 
This research was supported by grant 159/99-3 from the Israel 
Science Foundation.


\begin{thebibliography}{99}
\bibitem{Note} For a non-singular spacetime, e.g., that 
of a star, $r \in (0,\infty)$ maps into $x \in (0,\infty)$, while for
a spacetime with an event horizon, $r$ maps 
into $x \in (-\infty,\infty)$ 
($x$ being the so-called ``tortoise'' coordinate) 
\cite{Price,Ching}.
\end{thebibliography}
\end{document}